\documentclass[twocolumn,superscriptaddress,pra,aps]{revtex4-1}
\usepackage{amsfonts}
\usepackage{amssymb}
\usepackage{amsmath}
\usepackage{mathrsfs}
\usepackage{graphicx}
\usepackage{epsfig}
\usepackage{color,xcolor}

\begin{document}

\title{Tunable Aharonov-Bohm cages through anti-$\mathcal{PT}$-symmetric
imaginary couplings}
\author{S. M. Zhang}
\author{H. S. Xu}
\author{L. Jin}
\email{jinliang@nankai.edu.cn}
\affiliation{School of Physics, Nankai University, Tianjin 300071, China}

\begin{abstract}
The Aharonov-Bohm (AB) cage enables localized confinement with
nondiffractive propagation for arbitrary excitation. In this study, we
introduce an anti-parity-time (anti-$\mathcal{PT}$) symmetric imaginary
coupling in a generalized Creutz ladder to construct a non-Hermitian AB cage
with tunable flat-band energy. We investigate compact localized states and
complete localization dynamics, and show that non-Hermiticity affects the
localization probability distributions and increases the oscillation period
of the AB cage dynamics. Non-Hermitian engineering of the decoupled core of
the AB cage is the essential point in our proposal. Our approach is widely
applicable to a more general situation and can facilitate the manipulation
of localization in physics.
\end{abstract}

\maketitle

\section{Introduction}

Flat bands are a peculiar category of band structure that have fixed
energies independent of the crystal momentum \cite{LeykamAPX18,LeykamAPL18}.
They arise from destructive interference in many physical systems \cite%
{VicencioPRA16,RamachandranPRB17,MaimaitiPRB19,YuLPR20,FlachPRB21}, and
possess the unique property that the electron group velocity is exactly
zero. This property gives rise to various exotic phenomena in many-body
physics \cite{Tasaki,WuPRL07,MondainiPRB18}. In addition, the eigenstates of
the flat bands, known as compact localized states (CLSs), are completely
localized in one or more finite unit cells \cite%
{FlachEL14,YXXiaoPRL17,GneitingPRB18}. This prevents the wave
transport and makes it possible to trap and steer the propagation of light
on a given region \cite{ZGChenNanophotonics20,PobleteAPX21}. However, the
perfectly localized modes will be destroyed after the introduction of
nonlinearity \cite{PobleteAPX21}. And some of flat-band systems can exhibit
interesting transport \cite{LeykamPRA12,LeykamPRB13}; in particular, the
direction of transport can be well controlled according to the excitation
site or input phase in a rhombic lattice \cite{VicencioPRL22}. In addition,
the dispersionless nature of the flat-band systems make them guarantee a
greatly enhanced density of states, which amplifies the effects of
interactions. As a result, minimal repulsive on-site interactions can lead
to superconductivity \cite{HerreroNAT18,DeanSCI19,YoungNP20}. Moreover,
extensive theoretical works have demonstrated the topological properties of
nearly flat-band systems supporting unusual fractional topological phases
\cite{DelgadoPRL09,SarmaPRL11,XGWenPRL11,NeupertPRL11}. The intriguing
physics associated with flat bands has motivated their experimental
realization in various one-dimensional and two-dimensional settings,
including optical waveguides \cite%
{MukherjeePRL15,VicencioPRL15,MukherjeeOL17,MukherjeePRL18}, cold atomic
gases in optical lattices \cite{JoPRL12,TaieSA15,DrostNP17}, and
metamaterials \cite%
{MasumotoNJP12,NakataPRB12,KajiwaraPRB16,SlotNP17,WhittakerPRL18}. Recently,
research interest in flat bands continuously increases in optics \cite%
{LChenAP23}, condensed matters, and material science. In particular, the
flat-band localization in the Creutz ladder can be finely controlled in the
superradiance lattices \cite{WangPRL21}.

Stemming from the competition between geometry and magnetic fields, the
Aharonov-Bohm (AB) cage is a special flat-band system with its energy bands
fully flat. Any excitation in the system is completely confined to a certain
region. In the presence of a magnetic $\pi $-flux (i.e., half a flux
quantum) per plaquette, the quasi-one-dimensional (quasi-1D) rhombic lattice
\cite{LonghiOL14,MukherjeeOL15} and the two-dimensional dice lattice \cite%
{VidalPRL98,NaudPRL01,BerciouxPRA09,BerciouxPRA11,SMZhangPRB20} are
celebrated models that support the AB cages and have drawn much attention. Advanced fabrication techniques allow these models to be realized
in the laboratory. A total flux of $\pi $ within each plaquette of the
rhombic waveguide lattice has been obtained by introducing the effective
coupling between different orbital modes \cite{VicencioPRL22}, inserting an
auxiliary waveguide in each plaquette \cite{KremerNC20}, or injecting the
light beam with an orbital angular momentum \cite{SzameitLight20} The
importance of AB cage is reflected in many aspects \cite%
{LibertoPRA19,GligoricPRA20,ChangAPL21}, ranging from topological edge
states \cite{PelegriPRA19} to flat-band lasers \cite{LonghiOL19}. The
presence of nonlinearity or interactions in the Creutz ladder has attracted
much attention \cite{Hatsugai20,SilPRB22,WilsonPRL21}. Disorder on the
Creutz ladder with interparticle interactions induces exotic many-body
localization dynamics \cite{Orito21}, while the repulsive Hubbard
interaction causes repulsively bound pairs in the photonic Creutz-Hubbard
ladder \cite{Platero20}.

In recent decades, the non-Hermitian systems have been widely investigated both
theoretically and experimentally due to their intriguing properties \cite%
{LJinPRA09,YDChongPRL10,LonghiPRA10,RuterNP10}, including exceptional points
(EPs) \cite{KDingNRP22,LChenNN23}, unique light transport and wave
propagation \cite{LJinPRL18,HSXuPRA21,HSXuPRA23}, and exotic topology \cite%
{LJinPRB19,SMZhangPRA20,HCWuPRB20,HCWuPRB21}. As interest in non-Hermitian
systems continues to grow, proposals for flat bands have emerged in a large
number of non-Hermitian systems \cite%
{RamezaniPRA17,LeykamPRB17,QiPRL18,ZyuzinPRB18,GePR18,SMZhangPRA19,LJinPRA19,SMZhangPRR20,PHePRA21,MaimaitiPRB21,PXLuOE21,IEEE20,LeykamPRL17,LeykamOL13}%
. Non-Hermitian flat bands can exhibit polynomial power increase of
flat-band eigenstates \cite{RamezaniPRA17,GePR18}, which have no counterpart
in Hermitian cases. However, most of these studies have focused on the flat
bands created at the EPs \cite{IEEE20,LeykamPRL17,LeykamOL13}.

In this study, we propose an ingenious mechanism for fabricating a
non-Hermitian AB cage that is not necessarily at the EP of the non-Hermitian
lattice. This is distinct from previous non-Hermitian AB cages that are
formed exactly at the EP of the non-Hermitian lattice. We introduce an
approach to properly incorporate non-Hermiticity into the Hermitian AB cage
to keep all the bands flat, thus creating the non-Hermitian AB cage. We
demonstrate our proposal using a generalized cross-stitch lattice. The
anti-parity-time (anti-$\mathcal{PT}$) symmetric imaginary coupling between
two sublattices of the generalized cross-stitch lattice forms a
non-Hermitian Creutz ladder with a fully flat spectrum. In the
time-evolution dynamics, the excitation is fully confined inside the nearest
neighbor unit cells, and the intensity of confinement can exhibit constant,
oscillating, and exponential growth at different degrees of non-Hermiticity.
At weak imaginary coupling strength, the flat-band spectrum is entirely
real; at strong imaginary coupling strength, the flat-band spectrum is
entirely imaginary. The non-Hermiticity extends the period of oscillation
when the spectrum is real, and causes the probability distribution of the
confined excitation to pile up when the spectrum is imaginary.

The remainder of the paper is organized as follows. In Sec. II, we introduce
the generalized cross-stitch lattice, present the flat bands, and
demonstrate the confinement mechanism. In Sec. III, we propose the
non-Hermitian Creutz ladder by introducing dissipation-induced imaginary
couplings, and demonstrate the CLSs of the non-Hermitian AB cage not at the
EP. In Sec. IV, we analyze the typical localization dynamics in the
non-Hermitian cage. In Sec. V, we discuss the experimental
implementation of the proposed non-Hermitian AB cage in the coupled
resonator optical waveguides. Conclusions and discussions are summarized in
Sec. VI, respectively.

\section{Cross-stitch lattice}

\begin{figure}[tb]
\includegraphics[ bb=0 395 522 755, width=8.8 cm, clip]{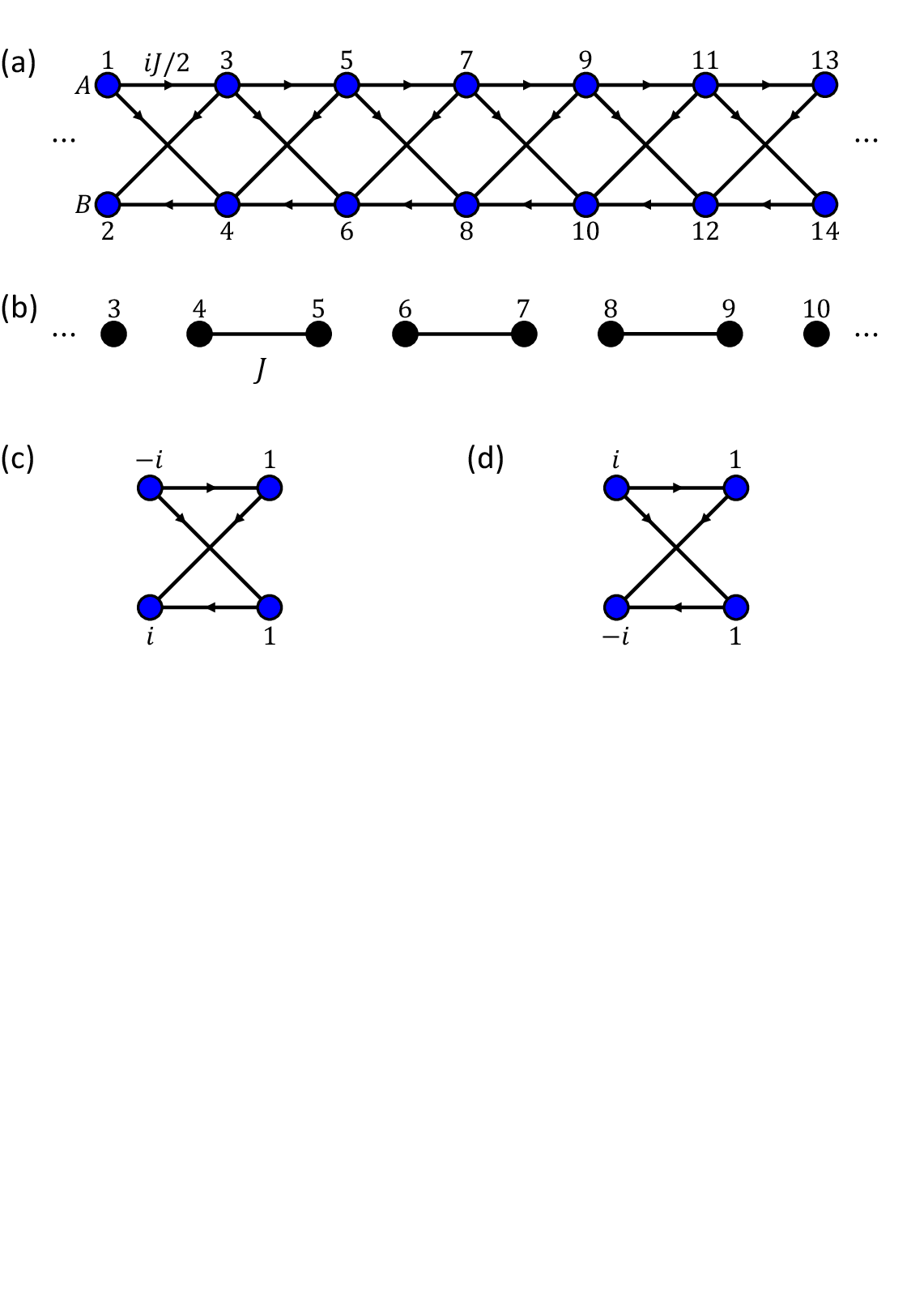}
\caption{(a) Schematic of the Hermitian cross-stitch lattice. (b) Equivalent lattice of decoupled dimers obtained by applying a similar transformation to (a). The reciprocal coupling strength is
$J$. (c) Compact localized eigenstate with the positive flat-band energy $J$. (d) Compact localized eigenstate with the negative flat-band energy $-J$.
The site numbers in (a) correspond to those in (b) under a similar
transformation.} \label{fig1}
\end{figure}

Figure~\ref{fig1}(a) schematically illustrates a generalization of the
conventional cross-stitch lattice \cite{Sharmaprr20}. Each coupling $\pm
iJ/2 $ has a nonreciprocal Peierls phase factor $e^{\pm i\pi /2}$. The sign $%
\pm $ of the nonreciprocal coupling $\pm iJ/2$ depends on the tunneling
direction of the particles. The particles tunneling along the black arrow
acquire a Peierls phase factor $e^{i\pi /2}$; while the particles tunneling
against the black arrow acquire a Peierls phase factor $e^{-i\pi /2}$. The
cross-stitch lattice can be viewed as sharing the same geometry as two
rhombic chains \cite{Platero20}, or alternatively, as the Creutz ladder
without rungs.

In spatial space, the Hamiltonian of the generalized cross-stitch lattice is
written in the form of%
\begin{eqnarray}
H &=&\sum_{n}[(iJ/2)b_{n}^{\dagger }(a_{n+1}+b_{n+1})  \notag \\
&&-(iJ/2)a_{n}^{\dagger }(a_{n+1}+b_{n+1})]+\mathrm{H.c.},
\end{eqnarray}%
where $a_{n}^{\dagger }$ ($a_{n}$) and $b_{n}^{\dagger }$ ($b_{n}$)
represent the creation (annihilation) operators of the sublattices $A$ and $%
B $, respectively. The lattice is translationally invariant in the
horizontal direction. The $j$-th unit cell of the generalized cross-stitch
lattice includes two sites $A_{j}$ and $B_{j}$. Notably, the generalized
cross-stitch lattice is Hermitian and does not include any intracell
coupling. In momentum space, the Bloch Hamiltonian under Fourier
transformation is written in the form of
\begin{equation}
H_{k}=(J\cos k)\sigma _{y}+(J\sin k)\sigma _{z},
\end{equation}%
where $\sigma _{x,y,z}$ is the Pauli matrix for the spin-$1/2$. The energy
bands $E_{k}=\pm J$ of $H_{k}$ are dispersionless and independent of the
momentum $k$. Therefore, the generalized cross-stitch lattice possesses an
entirely flat spectrum and supports an AB cage.

Now, we introduce an equivalent lattice as shown in Fig. \ref{fig1}(b) to
illustrate the kernel of the AB cage in the generalized cross-stitch
lattice. The schematic provides a concise and clear physical picture to
depict the essential feature of the generalized cross-stitch lattice. The
Hamiltonian $H^{\prime }$ of the equivalent lattice is connected to the
Hamiltonian $H$ of the generalized cross-stitch lattice under a similar
transformation $H^{\prime }=UHU^{-1}$. The most intriguing fact about the
equivalent lattice is the decoupled dimerization. This
generally originates from the distractive interference between the hoppings.
Figure~\ref{fig1}(b) illustrates the equivalent lattice $H^{\prime }$ of the
generalized cross-stitch lattice, which is fully constituted by the
decoupled dimers. Each dimer including a reciprocal coupling $J$ and all the
dimers are identical. Thus, the spectrum of the proposed AB cage is composed
of two flat bands $\pm J$ and the eigenstates are definitely localized in
each isolated dimer of the equivalent lattice. Notably, the equivalence to a
lattice of decoupled polymers is a feature for any AB cage from the fact
that all the bands are flat. From the equivalent lattice, we easily obtain
the confinement of the original AB cage. This also helps further engineering
the AB cage and the confinement.

As a result, the corresponding eigenstates of the generalized cross-stitch
lattice are also compactly localized. Both CLSs of the generalized
cross-stitch lattice in Fig. \ref{fig1}(a) are distributed and confined in a
four-site plaquette configuration. These CLSs are obtained from the
steady-state Schr\"{o}dinger equations. The CLS is $[-i,i,1,1]^{\mathrm{T}%
}/2 $ for the flat-band energy $J$ [Fig. \ref{fig1}(c)] and the CLS is $%
[i,-i,1,1]^{\mathrm{T}}$ for the flat-band energy $-J$ [Fig. \ref{fig1}(d)].
The localization of the CLSs is attributed to destructive interference.

The conventional cross-stitch lattice is similar to the generalized
configuration in Fig. \ref{fig1}(a), except that the phase factor $e^{\pm
i\pi /2}$ only sticks to the upper and lower chains. The cross couplings $%
J/2 $ between sublattices $A$ and $B$ maintain the two flat energy bands $%
\pm J$ \cite{ChanPRB22}, with both interchain and intrachain hopping
amplitudes set to $J/2$. The absence or presence of the phase factor $e^{\pm
i\pi /2}$ in the cross couplings is the only difference between the
conventional and generalized cross-stitch lattices. Both types of
cross-stitch lattices originate from the same equivalent lattice as shown in
Fig. \ref{fig1}(b). However, additional intracell couplings $%
\sum_{n}(e^{i\pi /2}\Omega a_{n}^{\dagger }b_{n}+e^{-i\pi /2}\Omega
b_{n}^{\dagger }a_{n})$ between sublattices in the vertical direction
significantly alter the flat bands. Adding intracell couplings to the
conventional cross-stitch lattice extends it into a four-flat-band Creutz
ladder, with the four flat-band energies being $\pm J\pm \Omega $ \cite%
{KunoPRB20}. By contrast, adding the vertical coupling $\Omega $ between
sublattices to the generalized cross-stitch lattice results in a two-band
Creutz ladder, with the two flat-band energies being $\pm \sqrt{J^{2}+\Omega
^{2}}$. The subtly designed phase factor in the cross couplings $\pm iJ/2$
of the generalized configuration in Fig. \ref{fig1}(a) plays a critical role
in maintaining the two-band structure. Imaginary (real) vertical couplings
decrease (increase) the band gap of the non-Hermitian (Hermitian) Creutz
ladder. In an interacting system, large repulsive interactions in the Creutz
ladder generate oscillating behavior in the local integrals of motion \cite%
{Orito21}.

\section{Non-Hermitian Creutz ladder}

\begin{figure}[t]
\includegraphics[ bb=0 33 522 391, width=8.8 cm, clip]{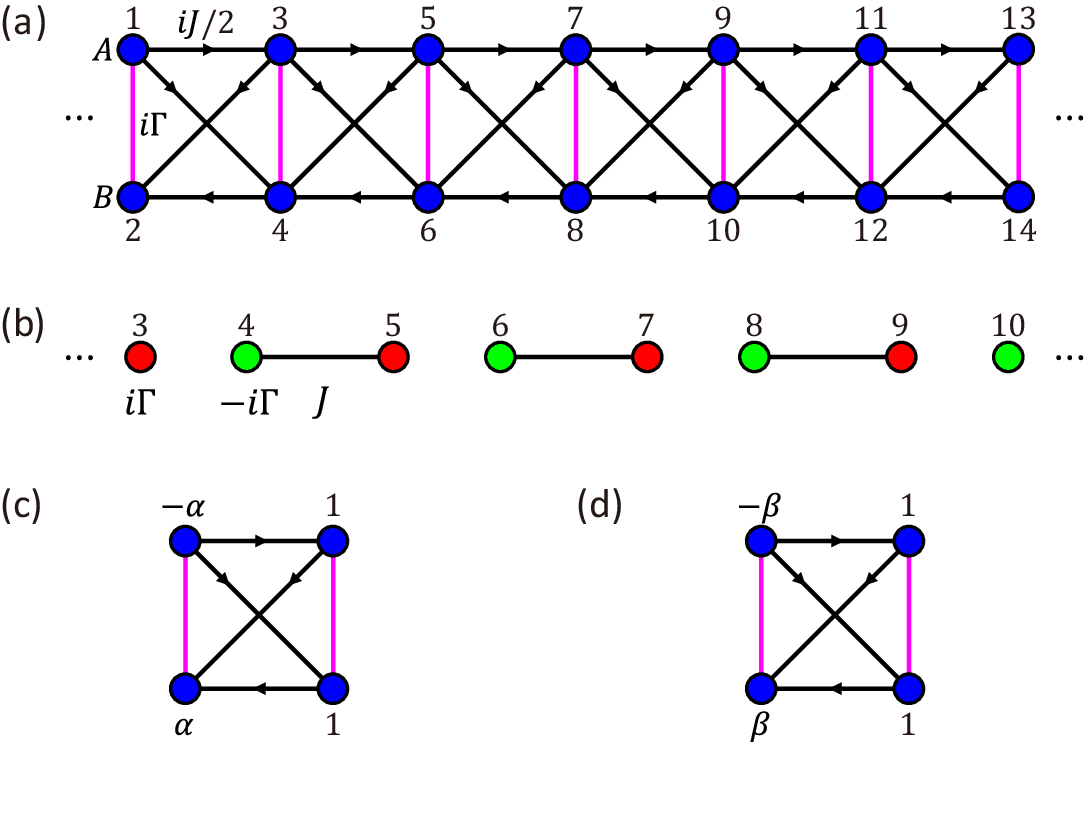}
\caption{(a) Schematic of the non-Hermitian Creutz ladder to construct the
AB cage not at the EP. Reciprocal non-Hermitian couplings $i{\Gamma }$
present in the vertical direction. (b) Schematic of the equivalent chain.
The non-Hermitian dimers are obtained by adding gain $+i\Gamma $ (red) and
loss $-i\Gamma $ (green) to Fig.~\protect\ref{fig1}(b). (c) Compact
localized eigenstate with the flat-band energy $\protect\sqrt{{J}^{2}-{\Gamma }^{2}}$. (d) Compact localized eigenstate with the flat-band energy $-\protect\sqrt{{J}^{2}-{\Gamma }^{2}}$. The signs in (c) and (d) are $\protect\alpha =(\Gamma -\protect\sqrt{\Gamma ^{2}-J^{2}})/J$, $\protect\beta =(\Gamma +\protect\sqrt{\Gamma ^{2}-J^{2}})/J.$}
\label{fig2}
\end{figure}

Figure~\ref{fig2}(a) shows the non-Hermitian Creutz ladder with imaginary
couplings introduced between two ladder legs \cite{Youpra17}. The intracell
couplings $i{\Gamma }$ are reciprocal, while the intercell couplings $\pm
iJ/2$ are nonreciprocal. The real-space Hamiltonian of the non-Hermitian
Creutz ladder reads
\begin{eqnarray}
\mathcal{H} &=&\sum_{n}-\frac{iJ}{2}a_{n}^{\dagger }(a_{n+1}+b_{n+1})+%
\mathrm{H.c.}  \notag \\
&&+\sum_{n}\frac{iJ}{2}b_{n}^{\dagger }(a_{n+1}+b_{n+1})+\mathrm{H.c.}
\notag \\
&&+\sum_{n}i\Gamma a_{n}^{\dagger }b_{n}+i\Gamma b_{n}^{\dagger }a_{n}.
\end{eqnarray}%
The imaginary coupling, satisfying the anti-$\mathcal{PT}$ symmetry \cite%
{HCWuPRB20,Konotop18,Gaonc22}, has been experimentally realized in various
systems such as the optical microcavity \cite{Wan20}, coupled waveguides
\cite{WWLiuLPR22}, optical fibres \cite{BergmanNC}, heat diffusion systems
\cite{Oius19}, cavity magnonics \cite{HuPRL20,DuJ20}, cold atoms \cite%
{WuJH16,Dus19}, electrical circuit resonators \cite%
{ChoiNC18,Aluprl21,Alunano21}, and quantum systems \cite{CZhengSR23}. The
flying atoms with ground state coherence indirectly account for the
imaginary couplings through coherently mixed spin waves \cite{XiaoN16}. Two
setups, one consisting of a spinning resonator driven by a pair of lasers
propagating in opposite directions \cite{JingNL20}, and the other comprising
a series of parallel cascaded resonators \cite{SunLPR23}, have already
induced the imaginary coupling. The auxiliary waveguides can be
adiabatically eliminated to realize the effective imaginary coupling in a
coupled waveguide array with alternately arranged auxiliary and primary
waveguides \cite{CTchan19,Sunprl22,LuOE19}. Additionally, a photonic system
consisting of two microresonators connected with two common waveguides is
accessible to obtain the dissipative coupling. The effective Hamiltonian of
this setup, expressed as a $2\times 2$ matrix, is reduced from the
Heisenberg-Langevin equations of microresonator modes. The indirect coupling
between microresonators indicated by nondiagonal terms of the effective
Hamiltonian is $-ie^{i\delta }\Gamma $, with the microresonator-waveguide
coupling $\Gamma $ and propagation phase factor of the probe light $%
e^{i\delta }$. When considering $\delta =\pm \pi $, the reciprocal indirect
coupling $-ie^{\pm i\pi }\Gamma $ is anti-$\mathcal{PT}$ symmetric and is
free of the gain-loss match in $\mathcal{PT}$ symmetry. Here, we concentrate
on the imaginary coupling $i\Gamma $.

\begin{figure*}[t]
\includegraphics[ bb=43 43 1255 582, width=18.0 cm, clip]{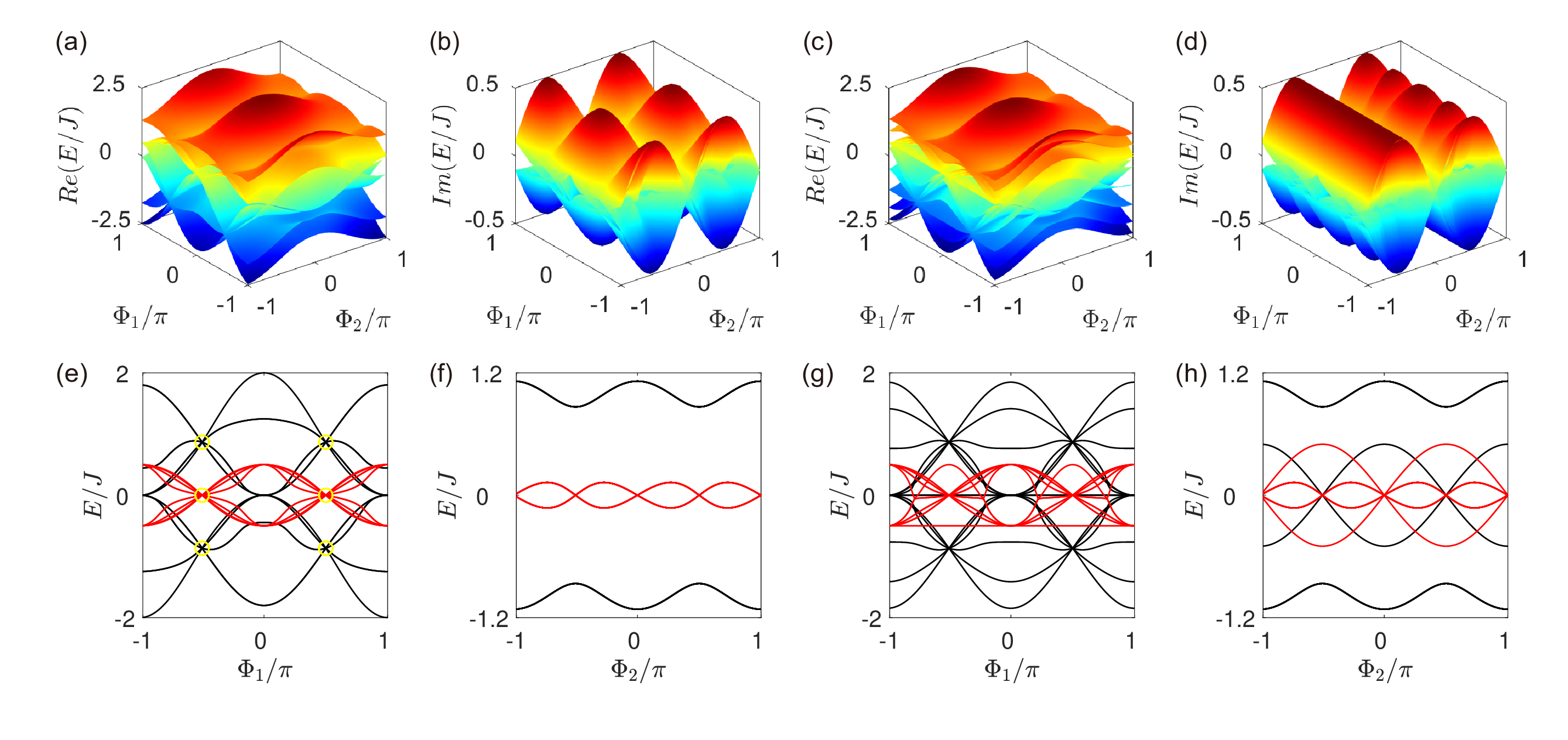}
\caption{Spectra of the real-space non-Hermitian Creutz ladder under
periodic boundary condition (left panel) and open boundary
condition (right panel). The phase $\Phi _{2}=\protect\pi /2$ is set in (e) and (g), causing a large number of energy levels coalesce to
two real energies indicated by yellow circles, where the AB cage forms. In (f), $\Phi _{1}=\protect\pi /2$, both bands are flat as $\Phi
_{2}$ changes. The real (imaginary) part in the four lower
panels is depicted in black (red). The Creutz ladder has $N=7$ unit cells,
with $J=1$ and $\Gamma =1/2$.}
\label{fig3}
\end{figure*}

Figure \ref{fig2}(b) shows an equivalent non-Hermitian dimerized chain.
Additional imaginary couplings $i{\Gamma }$ in Fig.~\ref{fig2}(a) yield
additional gain and loss $\pm i\Gamma $ in each isolated dimer of Fig.~\ref%
{fig2}(b). The non-Hermitian generalization maintains the AB cage, but
affects the flat-band energy and localization dynamics. We further elaborate
on the non-Hermitian isolation to proceed with our discussion. The gain and
loss $i\Gamma \sigma _{z}$ relate to the imaginary coupling $i\Gamma \sigma
_{x}$ through a similar transformation%
\begin{equation}
\left(
\begin{array}{cc}
i & i \\
-1 & 1%
\end{array}%
\right) ^{-1}\left(
\begin{array}{cc}
i\Gamma & 0 \\
0 & -i\Gamma%
\end{array}%
\right) \left(
\begin{array}{cc}
i & i \\
-1 & 1%
\end{array}%
\right) =\left(
\begin{array}{cc}
0 & i\Gamma \\
i\Gamma & 0%
\end{array}%
\right) .
\end{equation}%
The non-Hermitian Creutz ladder in Fig.~\ref{fig2}(a) is obtained by
applying the inverse similar transformation to the non-Hermitian dimerized
lattice in Fig.~\ref{fig2}(b). In this situation, the Bloch Hamiltonian of
the non-Hermitian Creutz ladder in momentum space reads
\begin{equation}
{\mathcal{H}_{k}}=\left(
\begin{array}{cc}
J\sin k & i\Gamma -iJ\cos k \\
i\Gamma +iJ\cos k & -J\sin k%
\end{array}%
\right) .
\end{equation}%
The band energies $E_{k}=\pm \sqrt{J^{2}-\Gamma ^{2}}$ of $\mathcal{H}_{k}$
are independent of the momentum $k$, and the imaginary coupling $\Gamma $
narrows the band gap. We notice that the non-Hermitian AB cage is not at the
EP except for $J=\Gamma $. This is a prominent difference from the
non-Hermitian AB cages proposed exactly at the EP, where the isolated dimers
are unidirectionally coupled \cite{IEEE20,SMZhangPRR20,LeykamPRB17}.
However, all the energy levels of the non-Hermitian Creutz ladder are
two-state coalesced at $J=\Gamma $.

Next, we discuss the eigenenergies and eigenstates of the non-Hermitian
Creutz ladder. We plot the eigenenergies of a $14$-site Creutz ladder in
Fig.~\ref{fig3}, where the nonreciprocal couplings $\pm iJ/2$ are
substituted by $e^{\pm i\Phi _{1}}J/2$ and the reciprocal couplings $i\Gamma
$ are substituted by $e^{i\Phi _{2}}\Gamma $. The AB cage forms at $\Phi
_{1}=\pi /2$ and is clearly exhibited in Fig.~\ref{fig3}(e). We
also show the wavefunction amplitudes of CLSs in Figs.~\ref{fig2}(c) and \ref%
{fig2}(d). The discrete Schr\"{o}dinger equations for the non-Hermitian
Creutz ladder are
\begin{eqnarray}
i\dot{\psi}_{A_{j}} &=&i\Gamma \psi _{B_{j}}-iJ/2\psi _{A_{j+1}}-iJ/2\psi
_{B_{j+1}}  \notag \\
&&+iJ/2\psi _{A_{j-1}}-iJ/2\psi _{B_{j-1}},  \label{A} \\
i\dot{\psi}_{B_{j}} &=&i\Gamma \psi _{A_{j}}+iJ/2\psi _{A_{j+1}}+iJ/2\psi
_{B_{j+1}}  \notag \\
&&+iJ/2\psi _{A_{j-1}}-iJ/2\psi _{B_{j-1}},  \label{B}
\end{eqnarray}%
where $\psi _{A_{j}}$ and $\psi _{B_{j}}$ are the wavefunctions at the $j$%
-th unit cell of the Creutz ladder. It is straightforward to verify that the
two CLSs shown in Figs.~\ref{fig2}(c) and \ref{fig2}(d) are the steady-state
solutions, and the wavefunctions outside the square plaquettes are zero.

\section{Non-Hermitian localization dynamics}

\begin{figure}[tb]
\includegraphics[ bb=0 0 600 712, width=8.8 cm, clip]{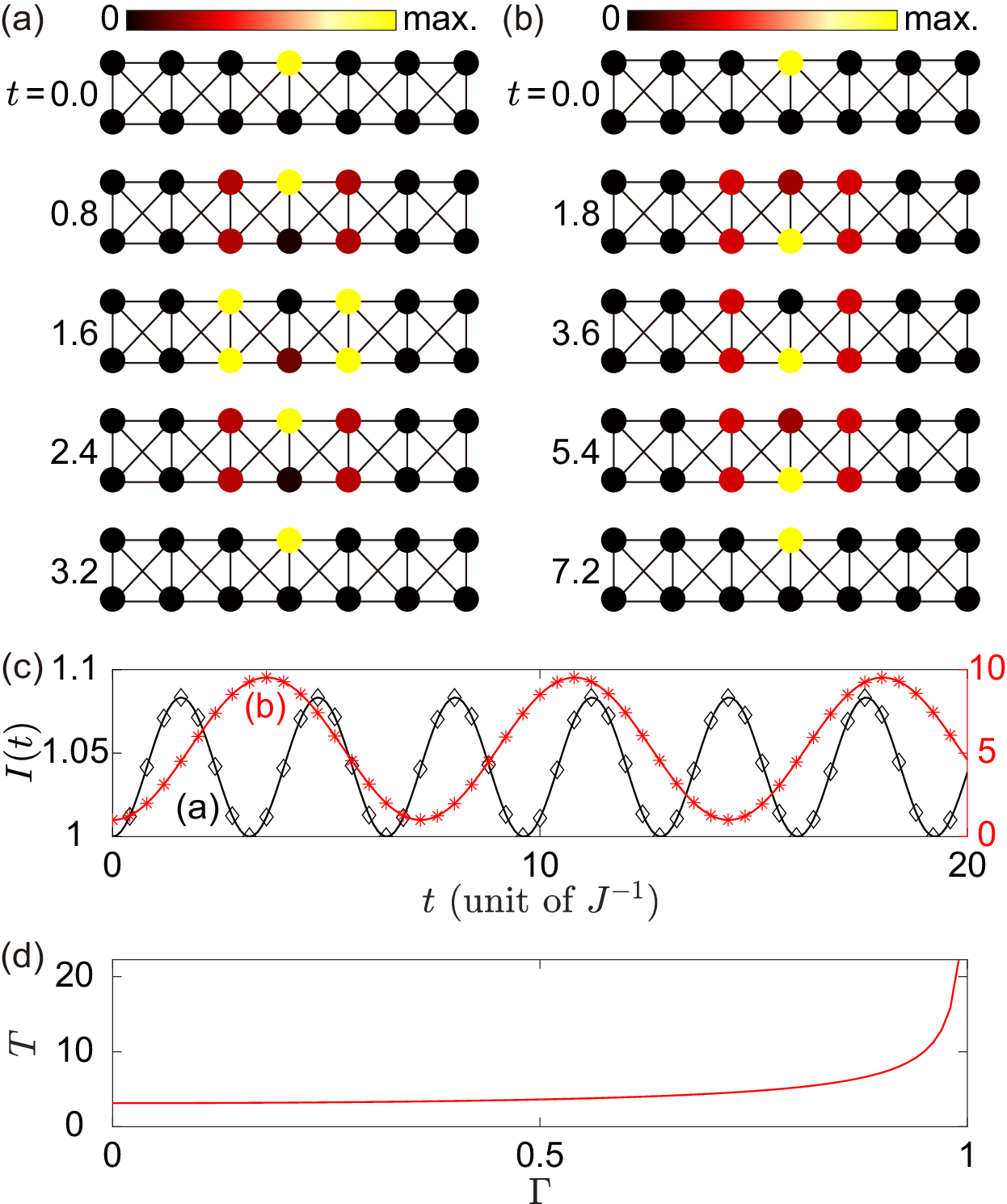}
\caption{(a, b) Localization dynamics with normalized intensity for
single-site excitations. The non-Hermitian Creutz ladder has $N=7$ unit
cells with $14$ sites. (c) Intensity of the time-evolution dynamics. The
initial excitation is at a single site shaded in yellow at $t=0$. The
parameters are fixed and being $J=1$, $\Gamma =1/5$ for (a); $J=1$, $\Gamma
=9/10$ for (b). The colored curves (marks) correspond to analytical
(numerical) results of intensities. (d) Periodicity $T$ with respect to the
non-Hermiticity $\Gamma $, where $T=\protect\pi /\protect\sqrt{J^{2}-\Gamma
^{2}}$ and $\Gamma \in \lbrack 0,1)$.} \label{fig4}
\end{figure}

The most significant and intriguing property of the non-Hermitian Creutz
ladder is the excitation confinement. The non-Hermitian localization
dynamics is solely induced by the intracell imaginary coupling, including
constant, oscillating, and increasing excitation intensities. Both bounded
and unbounded excitation intensities are compactly localized in finite sites
as a consequence of the localization of the CLSs. In this section, we
concentrate on the localization dynamics of the non-Hermitian Creutz ladder
for different excitations under open boundary condition.

We analyze the localization dynamics of a single-site excitation from an
analytical perspective. The excitation at sublattice $A$ and the excitation
at sublattice $B$ are two types of single-site excitations in the
non-Hermitian Creutz ladder. The localization dynamics is inevitable for any
excitation due to the fully flat spectrum. Any single-site excitation is
completely confined within six sites, as the initial state labeled $\psi
_{1}(0)=[0,0,1,0,0,0]^{\mathrm{T}}$\ can be expressed as a superposition of
four CLSs%
\begin{eqnarray}
\psi _{1}(0) &=&-\frac{\beta }{2\alpha -2\beta }[-\alpha ,\alpha ,1,1,0,0]^{%
\mathrm{T}}  \notag \\
&&+\frac{\alpha }{2\alpha -2\beta }[-\beta ,\beta ,1,1,0,0]^{\mathrm{T}}
\notag \\
&&-\frac{1}{2\alpha -2\beta }[0,0,-\alpha ,\alpha ,1,1]^{\mathrm{T}}  \notag
\\
&&+\frac{1}{2\alpha -2\beta }[0,0,-\beta ,\beta ,1,1]^{\mathrm{T}}.
\end{eqnarray}%
Similarly, the superposition coefficients of the initial state $\psi
_{2}(0)=[0,0,0,1,0,0]^{\mathrm{T}}$\ are given in the form of
\begin{eqnarray}
\psi _{2}(0) &=&-\frac{\beta }{2\alpha -2\beta }[-\alpha ,\alpha ,1,1,0,0]^{%
\mathrm{T}}  \notag \\
&&+\frac{\alpha }{2\alpha -2\beta }[-\beta ,\beta ,1,1,0,0]^{\mathrm{T}}
\notag \\
&&+\frac{1}{2\alpha -2\beta }[0,0,-\alpha ,\alpha ,1,1]^{\mathrm{T}}  \notag
\\
&&-\frac{1}{2\alpha -2\beta }[0,0,-\beta ,\beta ,1,1]^{\mathrm{T}}.
\end{eqnarray}

\begin{figure}[tb]
\includegraphics[ bb=0 37 597 255, width=8.8 cm, clip]{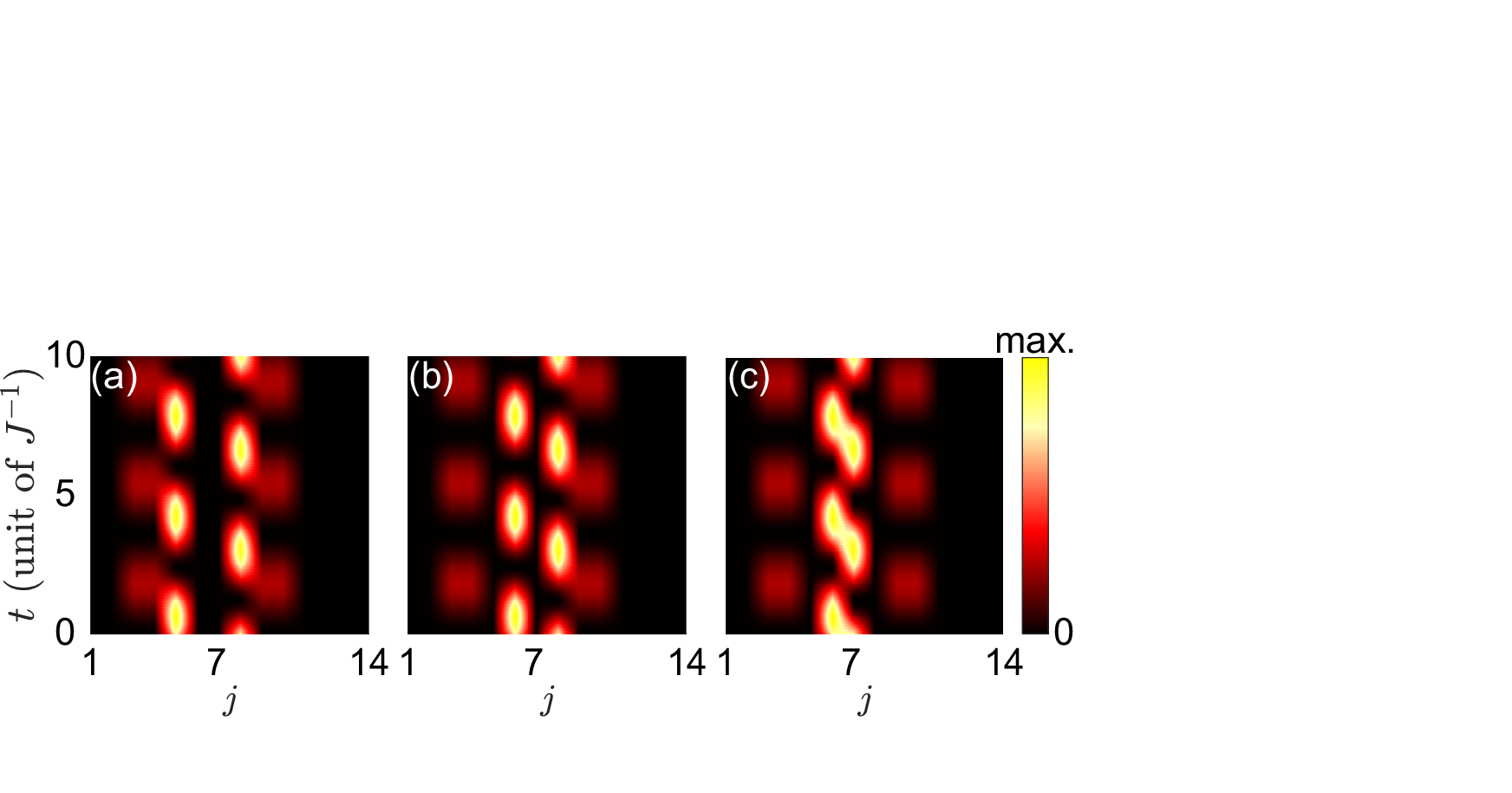}
\caption{The excitation confinement in the non-Hermitian Creutz ladder for
the real spectrum at $J>\Gamma $. (a) The initial state is $[0,0,0,0,1,0,0,-1,0,0,0,0,0,0]^{\mathrm{T}}/\protect\sqrt{2}$. (b) The initial state is $[0,0,0,0,0,1,0,1,0,0,0,0,0,0]^{\mathrm{T}}/\protect\sqrt{2}$. (c) The initial state is $[0,0,0,0,0,1,1,0,0,0,0,0,0,0]^{\mathrm{T}}/\protect\sqrt{2}$. The system
parameters are $J=1$, $\Gamma =1/2$, and $N=7$.} \label{fig5}
\end{figure}

\begin{figure}[tb]
\includegraphics[ bb=0 37 597 255, width=8.8 cm, clip]{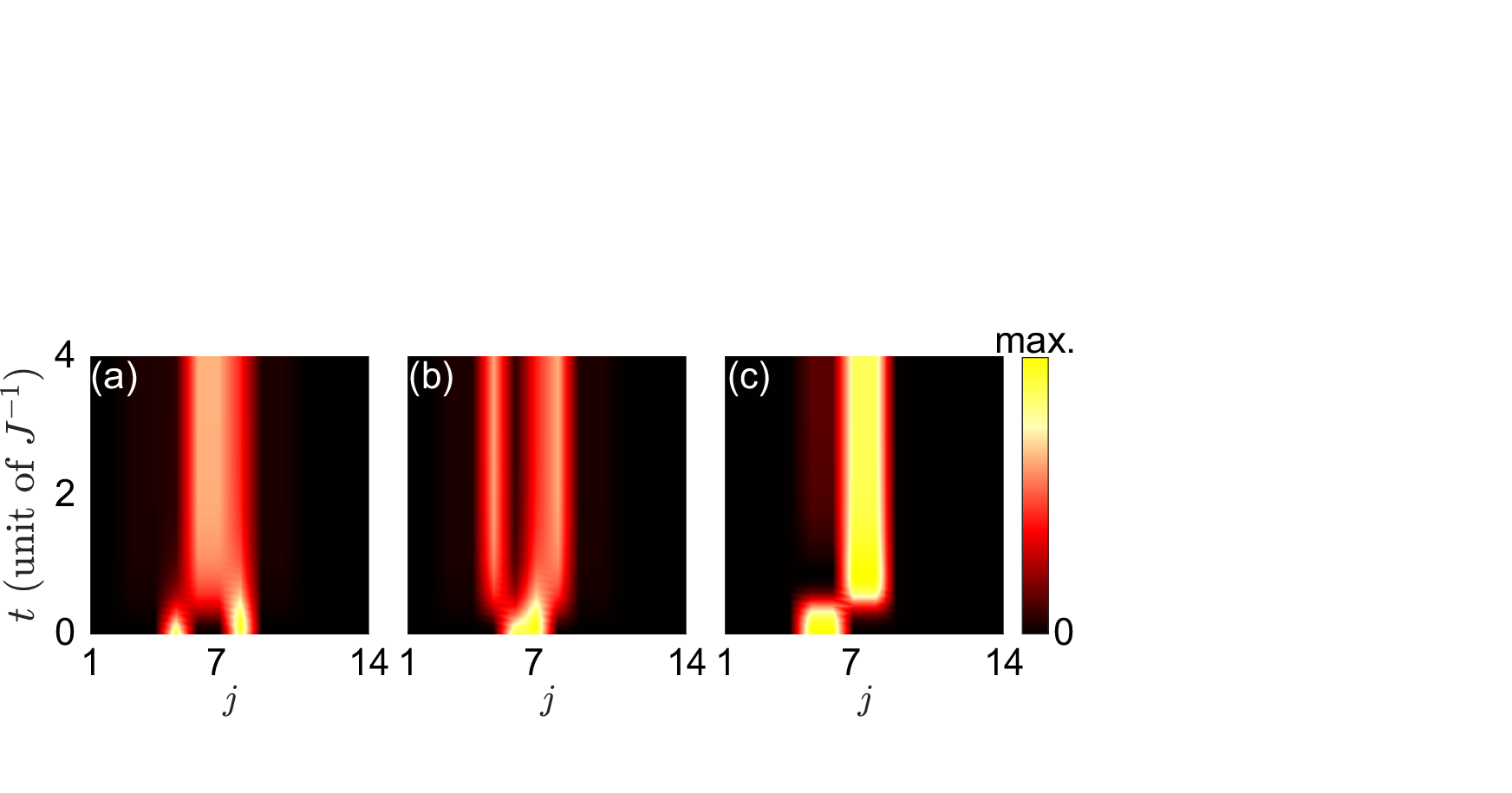}
\caption{The excitation confinement in the non-Hermitian Creutz ladder for
the imaginary spectrum at $J<\Gamma $. The total intensity is normalized to
unity. (a) The initial state is $[0,0,0,0,1,0,0,1,0,0,0,0,0,0]^{\mathrm{T}}/\protect\sqrt{2}$. (b) The initial state is $[0,0,0,0,0,1,-1,0,0,0,0,0,0,0]^{\mathrm{T}}/\protect\sqrt{2}$.
(c) The initial state is $[0,0,0,0,1,-1,0,0,0,0,0,0,0,0]^{\mathrm{T}}/\protect\sqrt{2}$. The system parameters are $J=1$, $\Gamma =3/2$, and $N=7$.}
\label{fig6}
\end{figure}

For a single-site excitation at the upper leg of the Creutz ladder, the
initial state is $\psi _{1}(0)$. Thus, the time-evolution state is
\begin{eqnarray}
\psi _{1}(t) &=&e^{-i\mathcal{H}t}\psi _{1}(0)  \notag \\
&=&[\frac{iJ}{2\omega }\sin \omega t,-\frac{iJ}{2\omega }\sin \omega t,\cos
\omega t,  \notag \\
&&-\frac{i\Gamma }{\omega }\sin \omega t,-\frac{iJ}{2\omega }\sin \omega t,-%
\frac{iJ}{2\omega }\sin \omega t]^{\mathrm{T}},  \label{C}
\end{eqnarray}%
where we set $\omega =\sqrt{J^{2}-\Gamma ^{2}}$ for simplicity. For a
single-site excitation at the lower leg of the Creutz ladder, the initial
state is $\psi _{2}(0)$, and the time-evolution state is
\begin{eqnarray}
\psi _{2}(t) &=&e^{-i\mathcal{H}t}\psi _{2}(0)  \notag \\
&=&[\frac{iJ}{2\omega }\sin \omega t,-\frac{iJ}{2\omega }\sin \omega t,-%
\frac{i\Gamma }{\omega }\sin \omega t,  \notag \\
&&\cos \omega t,\frac{iJ}{2\omega }\sin \omega t,\frac{iJ}{2\omega }\sin
\omega t]^{\mathrm{T}}.  \label{D}
\end{eqnarray}%
Figures~\ref{fig4} and~\ref{fig5} depict the bounded intensity for the real
spectrum at $J>\Gamma $. In contrast, in Fig.~\ref{fig6}, the intensity is
unbounded and increases with time for the imaginary spectrum at $J<\Gamma $.
Moreover, the time-evolution dynamics for any initial excitation can be
analytically obtained from the superposition of these two types of
single-site excitation dynamics.

In the following, we discuss the localization dynamics in two different
phases in detail. In the case of $J>\Gamma $ with a fully real flat
spectrum, the remarkable feature of the non-Hermitian Creutz ladder is the
oscillation period affected by $\Gamma $ in the time-evolution process.
Figure \ref{fig4} performs numerical simulations of two types of single-site
excitations, where the dynamics for the non-Hermitian Creutz ladder with the
oscillation period $\pi /\sqrt{J^{2}-\Gamma ^{2}}$ can be observed. The
non-Hermiticity plays a key role in extending the dynamic cycle. Moreover,
the peculiar fusion phenomenon is depicted in Fig.~\ref{fig5} by changing
the initial states, and the localization area ranges from sites $3$ to $10$
in all three panels.

In the case of $J<\Gamma $ with a fully imaginary spectrum, the
non-Hermitian Creutz ladder exhibits a prominently surging intensity in the
time-evolution dynamics. The confinement still remains because of the
flatness of energy bands although the spectrum is not real. In Fig. \ref%
{fig6}, the normalized intensity is skillfully depicted to demonstrate the
amplitudes, which eliminates the effect of sharply increased intensity and
helps to clearly observe the dynamic change of wavefunction amplitudes.
Figure~\ref{fig6} shows the exotic phenomena of convergence, splitting and
transfer during the evolution. The analytical time-evolution dynamics in
Eqs. (\ref{C})-(\ref{D}) helps to obtain the results in Fig. \ref{fig6}.
Since the time-evolution processes only involve several middle sites, the
boundary condition has no influence on the localization behaviors. However,
when considering an initial state excited at the lattice boundary,
intriguing phenomenon appears in the Creutz ladder under open boundary
condition.

Under open boundary condition, the dynamic behavior concerning the edge
state of the non-Hermitian Creutz ladder is also noteworthy, where the
intensity may exponentially decay with time. This image, resulting from the
boundary condition, breaks the inherent cognition that intensity increases
with time in the time-evolution process. The normalized right edge state is $%
\left\vert \psi _{R}\right\rangle =[0,0,\cdots ,1,-1]^{\mathrm{T}}/\sqrt{2}$
with the energy $-i\Gamma $ under open boundary condition, which is
topologically protected due to the topological equivalence of the two
lattices in Fig.~\ref{fig2}. The time-evolution state $\left\vert \psi
(t)\right\rangle $ for the initial state $\left\vert \psi _{R}\right\rangle $
is directly obtained, and expressed as $\left\vert \psi (t)\right\rangle
=e^{-i\mathcal{H}t}\left\vert \psi _{R}\right\rangle =e^{-\Gamma
t}[0,0,\cdots ,1,-1]^{\mathrm{T}}/\sqrt{2}$. The time-evolution dynamics
indicates that wavefunction amplitudes only distribute on the right
boundary. The corresponding intensity of the $\left\vert \psi
(t)\right\rangle $ is $I(t)=e^{-2\Gamma t}$, which decreases sharply with
time due to its exponential form. Moreover, the nonzero wavefunction
amplitudes only distribute for a short time, and the light propagation is
interrupted under open boundary condition. As for the left edge state $%
\left\vert \psi _{L}\right\rangle =[1,1,\cdots ,0,0]^{\mathrm{T}}/\sqrt{2}$\
with energy $i\Gamma $, the exponentially increasing intensity of the
time-evolution state is $I^{\prime }(t)=e^{2\Gamma t}$. As a comparison,
intensities for two edge states at the dissipative coupling $-i\Gamma $\
have opposite results under open boundary condition. The decaying intensity $%
e^{-2\Gamma t}$\ corresponds to the left edge state $\left\vert \psi
_{L}\right\rangle $\ with energy $-i\Gamma $, while the increasing intensity
$e^{2\Gamma t}$\ corresponds to the right edge state $\left\vert \psi
_{R}\right\rangle $\ with energy $i\Gamma $.

\section{Experimental realization}
The coupled resonator optical waveguides is a prominent platform for the
realization of discrete lattice models and many intriguing phenomena in
physics \cite{HSXuPRR22,LCXieSB23}. The coupled resonator optical waveguides
are comprised of a sequence of ring resonators, where the high precision
modulation and manipulation of the light field are possible. Here, we
propose the non-Hermitian Creutz ladder using the coupled resonator optical
waveguides.

\begin{figure}[tb]
\includegraphics[bb=0 0 521 274, width=8.8 cm, clip]{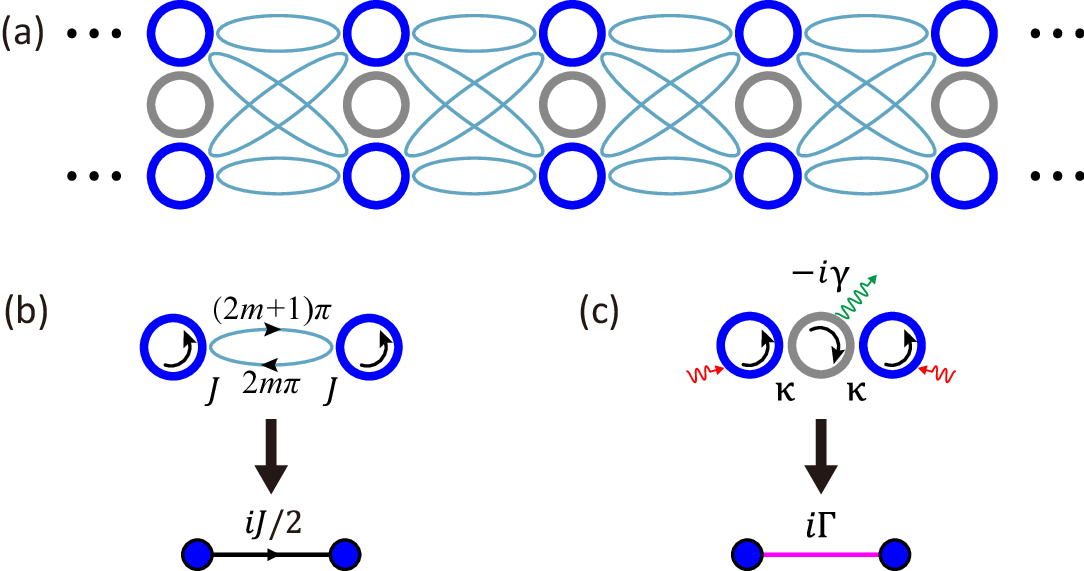}
\caption{(a) Schematic diagram of the coupled resonator optical waveguide. The primary resonators (blue rings) are
coupled to each other via the connecting waveguides (cyan ellipses) and
linking resonators (gray rings). (b) The realization of nonreciprocal
Hermitian coupling $\pm iJ/2$, where photons tunneling between the resonators
through the upper and lower paths along the connecting waveguide cumulate a phase factor of $e^{i(2m+1)\pi}$ and $e^{2m\pi}$, respectively. (c) The
realization of reciprocal non-Hermitian coupling $i\Gamma$. Two primary
resonators have the gain $i\Gamma$, the linking resonator has the loss $i\gamma$, the primary resonators and the linking resonator are coupled at the coupling strength $\kappa$.}
\label{fig7}
\end{figure}

Figure~\ref{fig7}(a) shows the schematic of a quasi-1D coupled ring
resonator array, where the blue rings are the primary resonators, the cyan
ellipses are the connecting waveguides, and the gray rings are the linking
resonators. The primary resonators are indirectly coupled through the
connecting waveguides and the linking resonators; and the primary resonators
on the upper and lower rows represent the sublattices $A$ and $B$,
respectively. The ring resonators support two degenerate clockwise and
counterclockwise modes. The clockwise mode and the counterclockwise mode are
decoupled. Notably, the mode chirality in the primary resonators and the
mode chirality in the linking resonators are opposite.

Each connecting waveguide induces a nonreciprocal coupling $\pm iJ/2$ as
shown in Fig.~\ref{fig7}(b). The coupling is Hermitian and has a
direction-dependent phase factor. The total length of the connecting
waveguide are designed to allow the accumulation of a phase shift of $%
e^{i(4m+1)\pi }$ when light propagates through it after a circle, where $m$
is an integer. Such design enables the constructive (destructive)
interference of photons within the primary resonators (connecting
waveguide). Consequently, the photons are confined in the primary resonators
rather than the waveguides \cite{HafeziNP11}. Furthermore, the lengths of
the upper and lower branches of the connecting waveguide are different. The
length difference causes a nonreciprocal phase. We consider all the primary
resonators in the system are coupled to the connecting waveguides with
identical coupling strength $J$. When a photon hops from the left resonator
to the right resonator, it accumulates a different phase $\pi $ than when
the photon hops in the opposite direction. Thus, the nonreciprocal hopping
phase factor is $e^{i\pi /2}$ for photons tunneling from left to right and
is $e^{-i\pi /2}$ for photons tunneling from right to left. Thus, the
primary resonators indirectly coupled through the connecting waveguides have
the left to right coupling $+iJ/2$ and the right to left $-iJ/2$ according
to coupled mode theory. The dynamics in the proposed coupled resonator
optical waveguides are governed by the Hamiltonian of the Hermitian
cross-stitch lattice.

To introduce the reciprocal non-Hermitian coupling $i\Gamma $, we use a
linking resonator illustrated in Fig.~\ref{fig7}(c) \cite{PXLuOE21}. The
linking resonators are designed to be on resonant with the primary
resonators; and the linking resonators are evanescently coupled to the
primary resonators with identical coupling strength $\kappa $. In the
situation that the linking resonator is lossy with an attenuation rate $%
\gamma \gg \kappa $, the large dissipation enables the adiabatical
elimination of the linking resonator light field and results in an effective
coupling $i\Gamma $ between the primary resonators at the strength $\Gamma
=\kappa ^{2}/\gamma $. However, the common loss $-i\Gamma $ also appear on
the two primary resonators after the adiabatic elimination process.
Therefore, the extra gain $i\Gamma $ on the two primary resonators are
required to balance the common loss $-i\Gamma $ induced by the linking
resonator. The dynamics in the proposed coupled resonator optical waveguides
are governed by the Hamiltonian of the non-Hermitian Creutz ladder.

\section{Conclusion and Discussion}

In this work, we propose an approach to construct the non-Hermitian AB cage
through destructive interference, with the essential point being the
formation of isolated non-Hermitian unit cells. We exemplify this approach
by constructing a non-Hermitian Creutz ladder with dissipation-induced
imaginary couplings introduced to the generalized cross-stitch lattice. The
interplay between Hermiticity and non-Hermiticity affects the flat-band gap,
compact localized states, and excitation confinement. The proposed
non-Hermitian AB cage, not necessarily at the EP, enriches the dynamics of
non-Hermitian localization. We offer a new manner and make a
step forward in modulating the flat-band energy, manipulating the light
confinement, and steering the period of localization. Our approach to
construct the non-Hermitian AB cage could be applicable for some other
quasi-1D systems such as the rhombic lattice \cite%
{KremerNC20,SzameitLight20,VicencioPRL22}. The non-Hermitian AB cage with
desirable localization properties can be directly constructed by designing
the decoupled non-Hermitian unit cells. Although the construction of
non-Hermitian AB cages in two-dimensional and three-dimensional systems
remains an open question, it is possible to generalize our idea to
higher-dimensional systems. Alternatively, it is interesting to further
consider the influence of the skin effect \cite%
{CHLeeFOP23,CHLeePRB23,SLKePRA23,JBGongPRB23} and the nonlinear effect \cite%
{LeykamPRA12,LeykamPRB13}. Our results may have potential application for
novel optical device design in the future.

\section*{Acknowledgments}

We acknowledge the support of National Natural Science Foundation of China
(Grant No. 11975128).

S. M. Zhang and H. S. Xu contributed equally to this work.


\begin{thebibliography}{999}
\bibitem{LeykamAPX18} D. Leykam, A. Andreanov, and S. Flach, Artificial flat
band systems: from lattice models to experiments, Adv. Phys. X \textbf{3},
1473052 (2018).

\bibitem{LeykamAPL18} D. Leykam and S. Flach, Perspective: Photonic
flatbands, APL Photonics \textbf{3}, 070901 (2018).

\bibitem{VicencioPRA16} L. Morales-Inostroza and R. A. Vicencio, Simple
method to construct flat-band lattices, Phys. Rev. A \textbf{94}, 043831
(2016).

\bibitem{RamachandranPRB17} A. Ramachandran, A. Andreanov, and S. Flach,
Chiral flat bands: Existence, engineering, and stability, Phys. Rev. B
\textbf{96}, 161104(R) (2017).

\bibitem{MaimaitiPRB19} W. Maimaiti, S. Flach, and A. Andreanov, Universal $%
d=1$ flatband generator from compact localized states, Phys. Rev. B \textbf{%
99}, 125129 (2019).

\bibitem{YuLPR20} D. Yu, L. Yuan, and X. Chen, Isolated photonic flatband
with the effective magnetic flux in a synthetic space including the
frequency dimension, Laser Photonics Rev. \textbf{14}, 2000041 (2020).

\bibitem{FlachPRB21} W. Maimaiti, A. Andreanov, and S. Flach, Flat-band
generator in two dimensions, Phys. Rev. B \textbf{103}, 165116 (2021).

\bibitem{Tasaki} H. Tasaki, Ferromagnetism in the Hubbard models with
degenerate single-electron ground states, Phys. Rev. Lett. \textbf{69}, 1608
(1992).

\bibitem{WuPRL07} C. Wu, D. Bergman, L. Balents, and S. Das Sarma, Flat
Bands and Wigner Crystallization in the Honeycomb Optical Lattice, Phys.
Rev. Lett. \textbf{99}, 070401 (2007).

\bibitem{MondainiPRB18} R. Mondaini, G. G. Batrouni, and B. Gremaud, Pairing
and superconductivity in the flat band: Creutz lattice, Phys. Rev. B \textbf{%
98}, 155142 (2018).

\bibitem{FlachEL14} S. Flach, D. Leykam, J. D. Bodyfelt, P. Matthies, and A.
S. Desyatnikov, Detangling flat bands into Fano lattices, Europhys. Lett.
\textbf{105}, 30001 (2014).

\bibitem{YXXiaoPRL17} Y.-X. Xiao, G. Ma, Z.-Q. Zhang, and C. T. Chan,
Topological Subspace-Induced Bound State in the Continuum, Phys. Rev. Lett.
\textbf{118}, 166803 (2017).

\bibitem{GneitingPRB18} C. Gneiting, Z. Li, and F. Nori, Lifetime of
flatband states, Phys. Rev. B \textbf{98}, 134203 (2018).

\bibitem{ZGChenNanophotonics20} L. Tang, D. Song, S. Xia, S. Xia, J. Ma, W.
Yan, Y. Hu, J. Xu, D. Leykam, and Z. Chen, Photonic flat-band lattices and
unconventional light localization, Nanophotonics \textbf{9}, 1161 (2020).

\bibitem{PobleteAPX21} R. A. V. Poblete, Photonic flat band dynamics, Adv.
Phys.: X \textbf{6}, 1878057 (2021).

\bibitem{LeykamPRA12} D. Leykam, O. Bahat-Treidel, and A. S. Desyatnikov,
Pseudospin and nonlinear conical diffraction in Lieb lattices, Phys. Rev. A
\textbf{86}, 031805(R) (2012).

\bibitem{LeykamPRB13} D. Leykam, S. Flach, O. Bahat-Treidel, and A. S.
Desyatnikov, Flat band states: Disorder and nonlinearity, Phys. Rev. B
\textbf{88}, 224203 (2013).

\bibitem{VicencioPRL22} G. C\'{a}ceres-Aravena, D. Guzm\'{a}n-Silva, I.
Salinas, and R. A. Vicencio, Controlled Transport Based on Multiorbital
Aharonov-Bohm Photonic Caging, Phys. Rev. Lett. \textbf{128}, 256602 (2022).

\bibitem{HerreroNAT18} Y. Cao, V. Fatemi, S. Fang, K. Watanabe, T.
Taniguchi, E. Kaxiras, and P. Jarillo-Herrero, Unconventional
superconductivity in magic-angle graphene superlattices, Nature (London)
\textbf{556}, 43 (2018).

\bibitem{DeanSCI19} M. Yankowitz, S. Chen, H. Polshyn, Y. Zhang, K.
Watanabe, T. Taniguchi, D. Graf, A. F. Young, and C. R. Dean, Tuning
superconductivity in twisted bilayer graphene, Science \textbf{363}, 1059
(2019)

\bibitem{YoungNP20} L. Balents, C. R. Dean, D. K. Efetov, and A. F. Young,
Superconductivity and strong correlations in moir\'{e} flat bands, Nat.
Phys. \textbf{16}, 725 (2020).

\bibitem{DelgadoPRL09} A. Bermudez, D. Patan\`{e}, L. Amico, and M. A.
Martin-Delgado, Topology-Induced Anomalous Defect Production by Crossing a
Quantum Critical Point, Phys. Rev. Lett. \textbf{102}, 135702 (2009).

\bibitem{SarmaPRL11} K. Sun, Z. Gu, H. Katsura, and S. D. Sarma, Nearly
Flatbands with Nontrivial Topology, Phys. Rev. Lett. \textbf{106}, 236803
(2011).

\bibitem{XGWenPRL11} E. Tang, J.-W. Mei, and X.-G. Wen, High-Temperature
Fractional Quantum Hall States, Phys. Rev. Lett. \textbf{106}, 236802 (2011).

\bibitem{NeupertPRL11} T. Neupert, L. Santos, C. Chamon, and C. Mudry,
Fractional Quantum Hall States at Zero Magnetic Field, Phys. Rev. Lett.
\textbf{106}, 236804 (2011).

\bibitem{MukherjeePRL15} S. Mukherjee, A. Spracklen, D. Choudhury, N.
Goldman, P. \"{O}hberg, E. Andersson, and R. R. Thomson, Observation of a
Localized Flat-Band State in a Photonic Lieb Lattice, Phys. Rev. Lett.
\textbf{114}, 245504 (2015).

\bibitem{VicencioPRL15} R. A. Vicencio, C. Cantillano, L. Morales-Inostroza,
B. Real, C. Mej\'{\i}a-Cort\'{e}s, S. Weimann, A. Szameit, and M. I. Molina,
Observation of Localized States in Lieb Photonic Lattices, Phys. Rev. Lett.
\textbf{114}, 245503 (2015).

\bibitem{MukherjeeOL17} S. Mukherjee and R. R. Thomson, Observation of
robust flat band localization in driven photonic rhombic lattices, Opt.
Lett. \textbf{42}, 2243 (2017).

\bibitem{MukherjeePRL18} S. Mukherjee, M. D. Liberto, P. \"{O}hberg, R. R.
Thomson, and N. Goldman, Experimental Observation of Aharonov-Bohm Cages in
Photonic Lattices, Phys. Rev. Lett. \textbf{121}, 075502 (2018).

\bibitem{JoPRL12} G.-B. Jo, J. Guzman, C. K. Thomas, P. Hosur, A.
Vishwanath, and D. M. Stamper-Kurn, Ultracold Atoms in a Tunable Optical
Kagome Lattice, Phys. Rev. Lett. \textbf{108}, 045305 (2012).

\bibitem{TaieSA15} S. Taie, H. Ozawa, T. Ichinose, T. Nishio, S. Nakajima,
and Y. Takahashi, Coherent driving and freezing of bosonic matterwave in an
optical Lieb lattice, Sci. Adv. \textbf{1}, 1500854 (2015).

\bibitem{DrostNP17} R. Drost, T. Ojanen, A. Harju, and P. Liljeroth,
Topological states in engineered atomic lattices, Nat. Phys. \textbf{13},
668 (2017).

\bibitem{MasumotoNJP12} N. Masumoto, N. Y. Kim, T. Byrnes, K. Kusudo, A. L%
\"{o}ffler, S. H\"{o}fling, A. Forchel, and Y. Yamamoto, Exciton-pololariton
condensates with flat bands in a two-dimensional kagome lattice, New J.
Phys. \textbf{14} 065002 (2012).

\bibitem{NakataPRB12} Y. Nakata, T. Okada, T. Nakanishi, and M. Kitano,
Observation of flat band for terahertz spoof plasmons in a metallic kagom%
\'{e} lattice, Phys. Rev. B \textbf{85}, 205128 (2012).

\bibitem{KajiwaraPRB16} S. Kajiwara, Y. Urade, Y. Nakata, T. Nakanishi, and
M. Kitano, Observation of a nonradiative flat band for spoof surface
plasmons in a metallic Lieb lattice, Phys. Rev. B \textbf{93}, 075126 (2016).

\bibitem{SlotNP17} M. R. Slot, T. S. Gardenier, P. H. Jacobse, G. C. P. van
Miert, S. N. Kempkes, S. J. M. Zevenhuizen, C. M. Smith, D. Vanmaekelbergh,
and I. Swart, Experimental realization and characterization of an electronic
Lieb lattice, Nat. Phys. \textbf{13}, 672 (2017).

\bibitem{WhittakerPRL18} C. E. Whittaker, E. Cancellieri, P. M. Walker, D.
R. Gulevich, H. Schomerus, D. Vaitiekus, B. Royall, D. M. Whittaker, E.
Clarke, I. V. Iorsh, I. A. Shelykh, M. S. Skolnick, and D. N. Krizhanovskii,
Exciton Polaritons in a Two-Dimensional Lieb Lattice with Spin-Orbit
Coupling, Phys. Rev. Lett. \textbf{120}, 097401 (2018).

\bibitem{LChenAP23} B. C. Xu, B. Y. Xie, L. H. Xu, M. Deng, W. J. Chen, H.
Wei, F. L. Dong, J. Wang, C. W. Qiu, S. Zhang, and L. Chen, Topological
Landau-Zener nanophotonic circuits, Adv. Photon, \textbf{5}, 036005 (2023).

\bibitem{WangPRL21} Y. He, R. Mao, H. Cai, J.-X. Zhang, Y. Li, L. Yuan,
S.-Y. Zhu, and D.-W. Wang, Flat-Band Localization in Creutz Superradiance
Lattices, Phys. Rev. Lett. \textbf{126}, 103601 (2021).

\bibitem{LonghiOL14} S. Longhi, Aharonov-Bohm photonic cages in waveguide
and coupled resonator lattices by synthetic magnetic fields, Opt. Lett.
\textbf{39}, 5892 (2014).

\bibitem{MukherjeeOL15} S. Mukherjee and R. R. Thomson, Observation of
localized flat-band modes in a quasi-one-dimensional photonic rhombic
lattice, Opt. Lett. \textbf{40}, 5443 (2015).

\bibitem{VidalPRL98} J. Vidal, R. Mosseri, and B. Dou\c{c}ot, Aharonov-Bohm
Cages in Two-Dimensional Structures, Phys. Rev. Lett. \textbf{81}, 5888
(1998).

\bibitem{NaudPRL01} C. Naud, G. Faini, and D. Mailly, Aharonov-Bohm Cages in
2D Normal Metal Networks, Phys. Rev. Lett. \textbf{86}, 5104 (2001).

\bibitem{BerciouxPRA09} D. Bercioux, D. F. Urban, H. Grabert, and W. H\"{a}%
usler, Massless Dirac-Weyl fermions in a ${\mathscr{T}}_{3}$ optical
lattice, Phys. Rev. A \textbf{80}, 063603 (2009).

\bibitem{BerciouxPRA11} D. Bercioux, N. Goldman, and D. F. Urban,
Topology-induced phase transitions in quantum spin Hall lattices, Phys. Rev.
A \textbf{83}, 023609 (2011).

\bibitem{SMZhangPRB20} S. M. Zhang and L. Jin, Compact localized states and
localization dynamics in the dice lattice, Phys. Rev. B \textbf{102}, 054301
(2020).

\bibitem{KremerNC20} M. Kremer, I. Petrides, E. Meyer, M. Heinrich, O.
Zilberberg, and A. Szameit, A square-root topological insulator with
non-quantized indices realized with photonic Aharonov-Bohm cages, Nat.
Commun. \textbf{11}, 907 (2020).

\bibitem{SzameitLight20} C. J\"{o}rg, G. Queralt\'{o}, M. Kremer, G. Pelegr%
\'{\i}, J. Schulz, A. Szameit, G. v. Freymann, J. Mompart, and V. Ahufinger,
Artificial gauge field switching using orbital angular momentum modes in
optical waveguides, Light Sci. Appl. \textbf{9}, 150 (2020).

\bibitem{LibertoPRA19} M. D. Liberto, S. Mukherjee, and N. Goldman,
Nonlinear dynamics of Aharonov-Bohm cages, Phys. Rev. A \textbf{100}, 043829
(2019).

\bibitem{GligoricPRA20} G. Gligori\'{c}, D. Leykam, and A. Maluckov,
Influence of different disorder types on Aharonov-Bohm caging in the diamond
chain, Phys. Rev. A \textbf{101}, 023839 (2020).

\bibitem{ChangAPL21} N. Chang, S. Gundogdu, D. Leykam, D. G. Angelakis, S.
Kou, S. Flach, and A. Maluckov, Nonlinear Bloch wave dynamics in photonic
Aharonov-Bohm cages, APL Photon. \textbf{6}, 030801 (2021).

\bibitem{PelegriPRA19} G. Pelegr\'{\i}, A. M. Marques, R. G. Dias, A. J.
Daley, J. Mompart, and V. Ahufinger, Topological edge states and
Aharanov-Bohm caging with ultracold atoms carrying orbital angular momentum,
Phys. Rev. A \textbf{99}, 023613 (2019).

\bibitem{LonghiOL19} S. Longhi, Photonic flat-band laser, Opt. Lett. \textbf{%
44}, 287 (2019).

\bibitem{Hatsugai20} Y. Kuno, T. Mizoguchi, and Y. Hatsugai,
Interaction-induced doublons and embedded topological subspace in a complete
flat-band system, Phys. Rev. A \textbf{102}, 063325 (2020).

\bibitem{SilPRB22} A. Mukherjee, A. Nandy, S. Sil, and A. Chakrabarti,
Tailoring flat bands and topological phases in a multistrand Creutz network,
Phys. Rev. B \textbf{105}, 035428 (2022).

\bibitem{WilsonPRL21} J. S. C. Hung, J. H. Busnaina, C. W. S. Chang, A. M.
Vadiraj, I. Nsanzineza, E. Solano, H. Alaeian, E. Rico, and C. M. Wilson,
Quantum Simulation of the Bosonic Creutz Ladder with a Parametric Cavity,
Phys. Rev. Lett. \textbf{127}, 100503 (2021).

\bibitem{Orito21} T. Orito, Y. Kuno, and I. Ichinose, Interplay and
competition between disorder and flat band in an interacting Creutz ladder,
Phys. Rev. B \textbf{104}, 094202 (2021).

\bibitem{Platero20} J. Zurita, C. E. Creffield, and G. Platero, Topology and
Interactions in the Photonic Creutz and Creutz-Hubbard Ladders, Adv. Quantum
Technol. \textbf{3}, 1900105 (2020).

\bibitem{LJinPRA09} L. Jin and Z. Song, Solutions of $\mathcal{PT}$%
-symmetric tight-binding chain and its equivalent Hermitian counterpart,
Phys. Rev. A \textbf{80}, 052107 (2009).

\bibitem{YDChongPRL10} Y. D. Chong, L. Ge, H. Cao, and A. D. Stone, Coherent
Perfect Absorbers: Time-Reversed Lasers, Phys. Rev. Lett. \textbf{105},
053901 (2010).

\bibitem{LonghiPRA10} S. Longhi, $\mathcal{PT}$-symmetric laser absorber,
Phys. Rev. A \textbf{82}, 031801(R) (2010).

\bibitem{RuterNP10} C. E. R\"{u}ter, K. G. Makris, R. El-Ganainy, D. N.
Christodoulides, M. Segev, and D. Kip, Observation of parity-time symmetry
in optics, Nat. Phys. \textbf{6}, 192 (2010).

\bibitem{KDingNRP22} K. Ding, C. Fang, and G. Ma, Non-Hermitian topology and
exceptional-point geometries, Nat. Rev. Phys. \textbf{4}, 745 (2022).

\bibitem{LChenNN23} A. Li, H. Wei, M. Cotrufo, W. Chen, S. Mann, X. Ni, B.
Xu, J. Chen, J. Wang, S. Fan, C.-W. Qiu, A. Al\`{u}, and L. Chen,
Exceptional points and non-Hermitian photonics at the nanoscale, Nat.
Nanotech. (2023).

\bibitem{LJinPRL18} L. Jin and Z. Song, Incident Direction Independent Wave
Propagation and Unidirectional Lasing, Phys. Rev. Lett. \textbf{121}, 073901
(2018).

\bibitem{HSXuPRA21} H. S. Xu, L. Jin, Coupling-induced nonunitary and
unitary scatteirng in anti-$\mathcal{PT}$-symmetric non-Hermitian systems,
Phys. Rev. A \textbf{104}, 012218 (2021).

\bibitem{HSXuPRA23} H. S. Xu , L. C. Xie, and L. Jin, High-order spectral
singularity, Phys. Rev. A \textbf{107}, 062209 (2023).

\bibitem{LJinPRB19} L. Jin, Z. Song, Bulk-boundary correspondence in a
non-Hermitian system in one dimension with chiral inversion symmetry, Phys.
Rev. B 99, 081103(R) (2019).

\bibitem{SMZhangPRA20} S. M. Zhang, X. Z. Zhang, L. Jin, Z. Song, High-order
exceptional point in supersymmetric array, Phys. Rev. A \textbf{101}, 033820
(2020).

\bibitem{HCWuPRB20} H. C. Wu, X. M. Yang, L. Jin, Z. Song, Untying links
through anti-parity-time-symmetric coupling, Phys. Rev. B \textbf{102},
161101(R)(2020).

\bibitem{HCWuPRB21} H. C. Wu, L. Jin, Z. Song, Topology of an
anti-parity-time symmetric non-Hermitian Su-Schirieffer-Heeger model, Phys.
Rev. B \textbf{103}, 235110 (2021).

\bibitem{RamezaniPRA17} H. Ramezani, Non-Hermiticity-induced flat band,
Phys. Rev. A \textbf{96}, 011802(R) (2017).

\bibitem{LeykamPRB17} D. Leykam, S. Flach, and Y. D. Chong, Flat bands in
lattices with non-Hermitian coupling, Phys. Rev. B \textbf{96}, 064305
(2017).

\bibitem{QiPRL18} B. Qi, L. Zhang, and L. Ge, Defect States Emerging from a
Non-Hermitian Flat Band of Photonic Zero Modes, Phys. Rev. Lett. \textbf{120}%
, 093901 (2018).

\bibitem{ZyuzinPRB18} A. A. Zyuzin and A. Yu. Zyuzin, Flat band in
disorder-driven non-Hermitian Weyl semimetals, Phys. Rev. B \textbf{97},
041203(R) (2018).

\bibitem{GePR18} L. Ge, Non-Hermitian lattices with a flat band and
polynomial power increase, Photon. Res. \textbf{6}, A10 (2018).

\bibitem{SMZhangPRA19} S. M. Zhang, L. Jin, Flat band in two-dimensional
non-Hermitian optical lattice, Phys. Rev. A \textbf{100}, 043808 (2019).

\bibitem{LJinPRA19} L. Jin, Flat band induced by the interplay of synthetic
magnetic flux and non-Hermiticity, Phys. Rev. A \textbf{99}, 033810 (2019).

\bibitem{SMZhangPRR20} S. M. Zhang, L. Jin, Localization in non-Hermitian
asymmetric rhombic lattice, Phys. Rev. Research \textbf{2}, 033127 (2020).

\bibitem{PHePRA21} P. He, H.-T. Ding, and S.-L. Zhu, Geometry and
superfluidity of the flat band in a non-Hermitian optical lattice, Phys.
Rev. A \textbf{103}, 043329 (2021).

\bibitem{MaimaitiPRB21} W. Maimaiti and A. Andreanov, Non-Hermitian
flat-band generator in one dimension, Phys. Rev. B \textbf{104}, 035115
(2021).

\bibitem{PXLuOE21} L. Ding, Z. Lin, S. Ke, B. Wang, and P. Lu, Non-Hermitian
flat bands in rhombic microring resonator arrays, Opt. Express \textbf{29},
24373 (2021).

\bibitem{LeykamPRL17} D. Leykam, K. Y. Bliokh, C. Huang, Y. D. Chong, and F.
Nori, Edge Modes, Degeneracies, and Topological Numbers in Non-Hermitian
Systems, Phys. Rev. Lett. \textbf{118}, 040401 (2017).

\bibitem{LeykamOL13} D. Leykam, V. V. Konotop, and A. S. Desyatnikov,
Discrete vortex solitons and parity time symmetry, Opt. Lett. \textbf{38},
371 (2013).

\bibitem{IEEE20} S. Ke, D. Zhao, J. Fu, Q. Liao, B. Wang, and P. Lu,
Topological Edge Modes in Non-Hermitian Photonic Aharonov-Bohm Cages, IEEE
J. Sel. Top. Quantum Electron. \textbf{26}, 4401008 (2020).

\bibitem{Sharmaprr20} R. Nehra, D. S. Bhakuni, A. Ramachandran, and A.
Sharma, Flat bands and entanglement in the Kitaev ladder, Phys. Rev.
Research \textbf{2}, 013175 (2020).

\bibitem{ChanPRB22} S. M. Chan, B. Gr\'{e}maud, and G. G. Batrouni, Pairing
and superconductivity in quasi-one-dimensional flat-band systems: Creutz and
sawtooth lattices, Phys. Rev. B \textbf{105}, 024502 (2022).

\bibitem{KunoPRB20} Y. Kuno, Extended flat band, entanglement, and
topological properties in a Creutz ladder, Phys. Rev. B \textbf{101}, 184112
(2020).

\bibitem{Youpra17} F. Yang, Y.-C. Liu, and L. You, Anti-$\mathcal{PT}$
symmetry in dissipatively coupled optical systems, Phys. Rev. A \textbf{96},
053845 (2017).

\bibitem{Konotop18} V. V. Konotop and D. A. Zezyulin, Odd-Time Reversal $%
\mathcal{PT}$ Symmetry Induced by an Anti-$\mathcal{PT}$-Symmetric Medium,
Phys. Rev. Lett. \textbf{120}, 123902 (2018).

\bibitem{Gaonc22} Y. Yang, X. Xie, Y. Li, Z. Zhang, Y. Peng, C. Wang, E. Li,
Y. Li, H. Chen, and F. Gao, Radiative anti-parity-time plasmonics, Nat.
Commun. \textbf{13}, 7678 (2022).

\bibitem{Wan20} F. Zhang, Y. Feng, X. Chen, L. Ge, and W. Wan, Synthetic
Anti-PT Symmetry in a Single Microcavity, Phys. Rev. Lett. \textbf{124},
053901 (2020).

\bibitem{WWLiuLPR22} W. Liu, Y. Zhang, Z. Deng, J. Ye, K. Wang, B. Wang, D.
Gao, and P. Lu, On-chip chiral mode switching by encircling an exceptional
point in an anti-parity-time symmetric system, Laser Photonics Rev. \textbf{%
16}, 2100675 (2022).

\bibitem{BergmanNC} A. Bergman, R. Duggan, K. Sharma, M. Tur, A. Zadok, and
A. Al\`{u}, Observation of anti-parity-time-symmetry, phase transitions and
exceptional points in an optical fibre, Nat. Commun. \textbf{12}, 486 (2021).

\bibitem{Oius19} Y. Li, Y.-G. Peng, L. Han, M.-A. Miri, W. Li, M. Xiao,
X.-F. Zhu, J. Zhao, A. Al\`{u}, S. Fan, and C.-W. Qiu, Anti-parity-time
symmetry in diffusive systems, Science \textbf{364}, 170 (2019).

\bibitem{HuPRL20} Y. Yang, Y.-P. Wang, J. W. Rao, Y. S. Gui, B. M. Yao, W.
Lu, and C.-M. Hu, Unconventional Singularity in Anti-Parity-Time Symmetric
Cavity Magnonics, Phys. Rev. Lett. \textbf{125}, 147202 (2020).

\bibitem{DuJ20} J. Zhao, Y. Liu, L. Wu, C.-K. Duan, Y.-X. Liu, and J. Du,
Observation of Anti-$\mathcal{PT}$-Symmetry Phase Transition in the
Magnon-Cavity-Magnon Coupled System, Phys. Rev. Appl. \textbf{13}, 014053
(2020).

\bibitem{WuJH16} X. Wang and J.-H. Wu, Optical $\mathcal{PT}$-symmetry and $%
\mathcal{PT}$-antisymmetry in coherently driven atomic lattices, Opt.
Express \textbf{24}, 4289 (2016).

\bibitem{Dus19} Y. Jiang, Y. Mei, Y. Zuo, Y. Zhai, J. Li, J. Wen, and S, Du,
Anti-Parity-Time Symmetric Optical Four-Wave Mixing in Cold Atoms, Phys.
Rev. Lett. \textbf{123}, 193604 (2019).

\bibitem{ChoiNC18} Y. Choi, C. Hahn, J. W. Yoon, and S. H. Song, Observation
of an anti-PT-symmetric exceptional point and energy-difference conserving
dynamics in electrical circuit resonators, Nat. Commun. \textbf{9}, 2182
(2018).

\bibitem{Aluprl21} H. Li, A. Mekawy, and A. Al\`{u}, Gain-Free Parity-Time
Symmetry for Evanescent Fields, Phys. Rev. Lett. \textbf{127}, 014301 (2021).

\bibitem{Alunano21} Z. Xiao and A. Al\`{u}, Tailoring exceptional points in
a hybrid PT-symmetric and anti-PT-symmetric scattering system, Nanophotonics
\textbf{10}, 3723 (2021).

\bibitem{CZhengSR23} J. Akram and C. Zheng, Theoretical investigation of
dynamics and concurrence of entangled $\mathcal{PT}$ and anti-$\mathcal{PT}$
symmetric polarized photons, Sci. Rep. \textbf{13}, 8542 (2023).

\bibitem{XiaoN16} P. Peng, W. Cao, C. Shen, W. Qu, J. Wen, L. Jiang, and Y.
Xiao, Anti-parity-time symmetry with flying atoms, Nat. Phys. \textbf{12},
1139 (2016).

\bibitem{JingNL20} H. Zhang, R. Huang, S.-D. Zhang, Y. Li, C.-W. Qiu, F.
Nori, and H. Jing, Breaking Anti-PT Symmetry by Spinning a Resonator, Nano
Lett. \textbf{20}, 7594 (2020).

\bibitem{SunLPR23} J. Zhang, Z. Feng, and X. Sun, Realization of Bound
States in the Continuum in Anti-PT-Symmetric Optical Systems: A Proposal and
Analysis, Laser Photonics Rev. \textbf{17}, 2200079 (2023).

\bibitem{CTchan19} X.-L. Zhang, T. Jiang, and C. T. Chan, Dynamically
encircling an exceptional point in anti-parity-time symmetric systems:
asymmetric mode switching for symmetry-broken modes, Light Sci. Appl.
\textbf{8}, 88 (2019).

\bibitem{Sunprl22} Z. Feng and X. Sun, Harnessing Dynamical Encircling of an
Exceptional Point in Anti-$\mathcal{PT}$-Symmetric Integrated Photonic
Systems, Phys. Rev. Lett. \textbf{129}, 273601 (2022).

\bibitem{LuOE19} S. Ke, D. Zhao, J. Liu, Q. Liu, Q. Liao, B. Wang, and P.
Lu, Topological bound modes in anti-PT-symmetric optical waveguide arrays,
Opt. Express \textbf{27}, 13858 (2019).

\bibitem{HSXuPRR22} H. S. Xu and L. Jin, Coherent resonant transmission,
Phys. Rev. Research \textbf{4}, L032015 (2022).

\bibitem{LCXieSB23} L. Xie, L. Jin, Z. Song, Antihelical edge states in
twodimensional photonic topological metals, Sci. Bull. \textbf{68}, 255
(2023).

\bibitem{HafeziNP11} M. Hafezi, E. A. Demler, M. D. Lukin, and J. M. Taylor,
Robust optical delay lines with topological protection, Nat. Phys. \textbf{7}%
, 907 (2011).

\bibitem{CHLeeFOP23} R. Lin, T. Tai, L. Li, and C. H. Lee, Topological
non-Hermitian skin effect, Front. Phys. \textbf{18}, 53605 (2023).

\bibitem{SLKePRA23} S. L. Ke, W. T. Wen, D. Zhao, and Y. Wang, Floquet
engineering of the non-Hermitian skin effect in photonic waveguide arrays,
Phys. Rev. A \textbf{107}, 053508 (2023).

\bibitem{CHLeePRB23} F. Qin, Y. Ma, R. Z. Shen, and C. H. Lee, Universal
competitive spectral scaling from the critical non-Hermitian skin effect,
Phys. Rev. B \textbf{107}, 155430 (2023).

\bibitem{JBGongPRB23} W. W. Zhu and J. B. Gong, Photonic corner skin modes
in non-Hermitian photonic crystals, Phys. Rev. B \textbf{108}, 035406 (2023).
\end{thebibliography}
\end{document}